\documentstyle[12pt]{article}

\begin{document}
\begin{flushright}
hep-th/9407134 \\
preprint UT-684 \\
July, 1994 \\
\end{flushright}

\bigskip

\begin{center}
{\large\bf The Topological $CP^1$ Model and the Large-$N$ \\
Matrix Integral}
\end{center}

\bigskip

\begin{center}
Tohru Eguchi

\medskip

{\it Department of Physics

\medskip

Faculty of Science

\medskip

University of Tokyo

\medskip

Tokyo 113, Japan }

\bigskip

and

\bigskip

Sung-Kil Yang

\medskip

{\it Institute of Physics

\medskip

University of Tsukuba

\medskip

Ibaraki 305, Japan }

\end{center}

\bigskip

\bigskip

\begin{abstract}
We discuss the topological $CP^1$ model which consists of the holomorphic
maps from Riemann surfaces onto $CP^1$. We construct a large-$N$ matrix model
which reproduces precisely the partition function of the $CP^1$ model
at all genera of Riemann surfaces. The action of our matrix model has the
form ${\rm Tr}\, V(M)=
-2{\rm Tr}\, M(\log M -1)+2\sum t_{n,P}{\rm Tr}\, M^n(\log M-c_n)
+\sum 1/n\cdot t_{n-1,Q}{\rm Tr}\, M^n~(c_n=\sum_1^n 1/j )$
where $M$ is an $N\times N$ hermitian matrix and $t_{n,P}\,
(t_{n,Q}),~(n=0,1,2\cdots)$ are the coupling constants
of the $n$-th descendant of the puncture (K\"ahler) operator.
\end{abstract}

\newpage

The topological $CP^1$ model was introduced a few years ago by \cite{Wa,DW}
as a characteristic example of a topological $\sigma$ model coupled to the
two-dimensional gravity. As is well-known, the topological
$\sigma$-model (or topological A model) is a twisted version of
the supersymmetric non-linear $\sigma$-model with a target space which
has an almost complex structure \cite{Wb}. Its path-integral is
given by a sum over contributions from instantons which are holomorphic
maps from the (genus $g$) Riemann surface onto a target space
$K$. When the target space $K$ is $CP^1$, the theory possesses two observables
$P$ and $Q$ corresponding to the identity and the K\"ahler class of $CP^1$
and thus the theory becomes a two-primary model.
When the system is coupled to the two-dimensional gravity,
additional excitations, e.g. gravitational
descendants $\sigma_n(P),~\sigma_n(Q)~(n=1,2,\cdots)$ appear in the theory
(by convention $\sigma_0(P)=P,\sigma_0(Q)=Q$).
We denote the coupling constants associated with these observables
as $t_{n,P},~t_{n,Q}~(n=0,1,2,\cdots)$.

In the following we construct a matrix
model which reproduces exactly the partition function of the $CP^1$ model
in the large phase space ($t_{n,P}\neq 0,~t_{n,Q}\neq 0$) at all genera of the
Riemann surface. The potential of our matrix model has the form
\begin{eqnarray}
&&V(M)=-2M(\log M-1)+2\sum_{n=1}t_{n,P}M^n(\log M-c_n)
\nonumber \\
&&~~~~~~~~~~~~+\sum_{n=1}\frac{1}{n}t_{n-1,Q}M^n~,
{}~~~~~~~c_n=\sum_{j=1}^{n}\frac{1}{j}
\end{eqnarray}
and the partition function is given by
\begin{equation}
Z=\int dM \exp(N{\rm Tr}\, V(M))~.
\end{equation}
Here $M$ is an $N\times N$ hermitian matrix.
The characteristic feature of our model (1) is the appearance of the
logarithmic potentials indicative of the asymptotic freedom and the
non-vanishing
$\beta$ function. In the limit when all coupling constants vanish
our action is reduced to $V(M)=-2M(\log M-1)$
which is the form of the effective
potential of the $CP^1$ model when $M$ is replaced by the chiral
superfield $\Phi$ \cite{CV}. The extremum of the action $V'(\Phi)=0$
occurs at $\Phi^2=1$ which is
the basic relation of the $CP^1$ quantum cohomology \cite{Wa,V}.

Let us first describe some basic features of the $CP^1$ model at genus $g=0$.
By denoting the free-energy of the system as $F$ basic two-point functions of
the theory are defined as
\begin{eqnarray}
&&\langle PP \rangle =\frac{\partial^2 F}{\partial t_{0,P}^2}\equiv u~,
\label{eq:1}
\\
&&\langle PQ \rangle =\frac{\partial^2 F}{\partial t_{0,Q}\partial t_{0,P}}
\equiv v~.
\label{eq:2}
\end{eqnarray}

A crucial ingredient of the $CP^1$ model is the following (constitutive)
relation
\begin{equation}
\langle QQ \rangle =e^{\langle PP\rangle }=e^u~
\label{eq:3}
\end{equation}
which comes from the instanton analysis \cite{Wa}.
(\ref{eq:3}) is the formula which
distinguishes the $CP^1$ model from other two-primary models like the Ising
model. From (\ref{eq:1}),(\ref{eq:2}),(\ref{eq:3}) we obtain the flow
equations for the time variable
$t_{0,Q}$
\begin{eqnarray}
&&\frac{\partial u}{\partial t_{0,Q}}= v'~,
\label{eq:4}
\\
&&\frac{\partial v}{\partial t_{0,Q}}=(e^u)'~.
\label{eq:5}
\end{eqnarray}
Here $'$ denotes the derivative in $t_{0,P}$. The flow equations for the
descendant times $t_{n,P},~t_{n,Q}~(n=1,2\cdots)$ are obtained by using the
genus $g=0$ topological recursion relations \cite{Wa}
\begin{equation}
\langle \sigma_n(\Phi_{\alpha})XY\rangle=n\langle \sigma_{n-1}
(\Phi_{\alpha})P \rangle \langle QXY \rangle +
n\langle \sigma_{n-1}(\Phi_{\alpha})Q \rangle \langle PXY \rangle~,
\end{equation}
where $\Phi_{\alpha}$=$P$ or $Q$. $\partial u/\partial t_{1,P}$ is,
for instance, derived as
\begin{eqnarray}
\frac{\partial u}{\partial t_{1,P}} &=& \langle \sigma_1(P)PP\rangle
=\langle PP\rangle \langle QPP\rangle +\langle PQ\rangle \langle PPP\rangle
\nonumber \\
&=& uv'+vu'=(uv)'~.
\end{eqnarray}
Similarly
\begin{eqnarray}
\frac{\partial v}{\partial t_{1,P}} &=& \langle \sigma_1(P)PQ \rangle
=\langle PP \rangle \langle QPQ \rangle +\langle PQ \rangle \langle PPQ \rangle
\nonumber \\
&=& u(e^u)'+vv'=(\frac{1}{2}v^2+(u-1)e^u)'~.
\end{eqnarray}
For the sake of illustration we also present the flow equations for
the variables $t_{1,Q},\, t_{2,P}$ and $t_{2,Q}$
\begin{eqnarray}
&&\frac{\partial u}{\partial t_{1,Q}}=(\frac{1}{2}v^2+e^u)'~,~~~~~~~~~~~~~
\frac{\partial v}{\partial t_{1,Q}}=(ve^u)'~,
\label{eq:8}
\\
&&\frac{\partial u}{\partial t_{2,P}}=(uv^2+2(u-2)e^u)'~,~~
\frac{\partial v}{\partial t_{2,P}}=(\frac{1}{3}v^3+2v(u-1)e^u)'~,
\\
&&\frac{\partial u}{\partial t_{2,Q}}=(\frac{1}{3}v^3+2ve^u)'~,~~~~~~~~~~
\frac{\partial v}{\partial t_{2,Q}}=(v^2e^u+e^{2u})'~.
\end{eqnarray}

It is clear that one can write down flow equations for any of the
variables $t_{n,P},\, t_{n,Q}$ by using the recursion relations repeatedly.
By construction these flows mutually commute with each other and thus
they define an
integrable hierarchy.

It is easy to see that the $CP^1$ system is closely related to
the Toda lattice hierarchy \cite{D}.
In fact by combing (\ref{eq:4}) and (\ref{eq:5})
we obtain
\begin{equation}
\frac{\partial^2 u}{\partial t_{0,Q}^2}=(e^u)''
\label{eq:6}
\end{equation}
which may be regarded as the continuum (genus $g=0$) limit of the Toda lattice
equation
\begin{equation}
\frac{\partial^2 u_n}{\partial z \partial \bar{z}}=e^{u_{n+1}}+e^{u_{n-1}}
-2e^{u_{n}}
\end{equation}
if we identify $t_{0,P}=n/N$ ($N$ is the size of the lattice).
In (\ref{eq:6}) $t_{0,Q}\approx z=\bar{z}$ and
thus we have the one-dimensional Toda theory.
As we shall see in the following, our system can in fact be described by a
suitable generalization of the one-dimensional Toda lattice hierarchy.

Let us now introduce the Lax formalism for the Toda lattice (at genus $g=0$)
and see how one can reproduce the evolution equations of the $CP^1$ model.
We consider an operator
\begin{equation}
L=p+v+e^u p^{-1}~,
\end{equation}
where the momentum $p$ is the canonical conjugate of $t_{0,P}$. The Poisson
bracket of the Toda system is defined as \cite{TT}
\begin{equation}
\{A,B\}=p\Big(\frac{\partial A}{\partial p}\frac{\partial B}{\partial t_{0,P}}-
\frac{\partial B}{\partial p}\frac{\partial A}{\partial t_{0,P}}\Big)~.
\end{equation}
Time-evolution is generated by the Hamiltonians $(L^n)_+,~
n=1,2\cdots$, where $+$ means taking terms with non-negative powers of $p$.
By introducing the Toda-times $t_n$ we have
\begin{equation}
\frac{\partial L}{\partial t_n}=\{(L^n)_+,L\}~,~~~~~~~n=1,2,\cdots~.
\label{eq:7}
\end{equation}
By comparing (\ref{eq:7}) with the flow equations of the $CP^1$ model
we find that the $CP^1$-times $t_{n,Q} (n=0,1,2\cdots)$ are nothing but
the Toda-times
$t_{n+1} (n=0,1,2\cdots)$
\begin{equation}
t_{n,Q}=(n+1)t_{n+1}~,~~~~~~~~~~~n=0,1,2,\cdots~.
\end{equation}
For instance,
\begin{eqnarray}
\frac{\partial L}{\partial t_1} &=& \{L_+,L\}=\{p+v,p+v+e^u p^{-1}\}
\nonumber \\
&=& (e^u)'+v'e^u p^{-1}~.
\end{eqnarray}
Hence $\partial v/\partial t_1=(e^u)',~\partial u/\partial t_1= v'$
which coincide with eqs.(\ref{eq:5}),(\ref{eq:4}). Similarly
\begin{eqnarray}
\frac{\partial L}{\partial t_2} &=& \{(L^2)_+,L\}
=\{p^2+2vp+v^2+2e^u,p+v+e^u p^{-1}\}
\nonumber \\
&=& 2(ve^u)'+(v^2+2e^u)'e^u p^{-1}~.
\end{eqnarray}
Hence $\partial v/\partial t_2=2(v e^u)',~\partial u/\partial t_2
=2(v^2/2+e^u)'$ which agree with eq.(\ref{eq:8}).
Thus the generators of the
$t_{n,Q}$-time evolutions are identified as simple powers of the Lax operator
$L$.

If one attempts to describe the $t_{n,P}$-time evolutions using the Lax
formalism, one has to generalize the standard treatment of the Toda theory
and introduce the logarithm of the Lax operator $\log L$.
We introduce a series of the operators $L^n(\log L-c_n)$ with
$c_n=\sum_{j=1}^{n}1/j$ and consider an additional set of evolution
equations
\begin{equation}
\frac{\partial L}{\partial s_n}=\{\big(L^n(\log L-c_n)\big)_+,L\}~,
{}~~n=0,1,2,\cdots.
\label{eq:9}
\end{equation}
It is possible to show that only terms proportional to $p^0$ and $p^{-1}$
appear in the
right-hand-side of eq.(\ref{eq:9}) as in the case of (\ref{eq:7})
(for the treatment of the logarithm see below,
eq.(\ref{eq:10})) and hence (\ref{eq:9}) gives rise to a set of differential
equations for the coefficient functions $u$ and $v$.
Furthermore, the flows in the variables $s_n$ commute among themselves
and also with those of the variables $t_m$.

Then by comparing (\ref{eq:9}) with the $CP^1~t_{n,P}$-flow equations we find
\begin{equation}
\frac{1}{2}s_n=t_{n,P}~,~~~~~~~~~~~~~n=0,1,2,\cdots.
\label{eq:11}
\end{equation}
Let us check the $n=0$ case. We first rewrite the $\log L$ operator as
\begin{eqnarray}
\log L &=& \log(p+v+e^u p^{-1})
\nonumber \\
&=& \frac{1}{2}\log p(1+vp^{-1}+e^up^{-2})
+\frac{1}{2}\log p^{-1}e^u(1+ve^{-u}p+e^{-u}p^2)
\nonumber \\
&=& \frac{1}{2}u+\frac{1}{2}\log(1+vp^{-1}+e^up^{-2})
+\frac{1}{2}\log(1+ve^{-u}p+e^{-u}p^2).
\label{eq:10}
\end{eqnarray}
Taking the positive part simply drops the 2nd term in (\ref{eq:10}).
Then by direct calculation,
\begin{equation}
\frac{\partial L}{\partial s_0}=\{\big(\log L\big)_+,L\}
=\frac{1}{2}(v'+(e^u)'p^{-1})~.
\end{equation}
Hence
\begin{equation}
s_0=2t_{0,P}~.
\end{equation}
We can check other cases $n=1,2,\cdots$ of (\ref{eq:11}) with somewhat lengthy
calculations. The coefficients $c_n$ are determined to eliminate the
$\partial/\partial t_{n-1,Q}$ component in the action of the operator
$(L^n\log L)_+$ on $L$.
Our operators $L^n(\log L-c_n)$ have a special ``scaling'' property
\begin{equation}
\frac {d}{dL} L^n(\log L-c_n)=nL^{n-1}(\log L-c_{n-1})
\label{eq:18}
\end{equation}
which plays an important role
in what follows.
We summarize our identification of the $CP^1$ evolution equations
at the genus $g=0$ level,
\begin{eqnarray}
&&n\frac{\partial L}{\partial t_{n-1,Q}}=\{(L^n)_+,L\}~,
{}~~~~~~~~~~~~~~~~~~~~~n=1,2,\cdots ,
\label{eq:12}\\
&&\frac{1}{2}\frac{\partial L}{\partial t_{n,P}}
=\{\big(L^n(\log L-c_n)\big)_+,L\}~,~~~~~~~~~~~n=1,2,\cdots~.
\label{eq:13}
\end{eqnarray}

Before we turn to the subject of the matrix-model realization of the
hierarchy (\ref{eq:12}),(\ref{eq:13})
we must discuss its higher-genus corrections.
As we discussed elsewhere \cite{EYY}, it is straightforward to determine
corrections to the flow equations of the $CP^1$ model at each genus simply
by demanding the commutativity of the flows. Up to the order of $g=2$ the
corrected flow equations (for $t_{0,Q}$, $t_{1,P}$ and $t_{1,Q}$) are given by
\begin{eqnarray}
\frac{\partial u}{\partial t_{0,Q}} &=& v'~,
\\
\frac{\partial v}{\partial t_{0,Q}} &=& \Big(e^u+\lambda^2\big(\frac{1}{24}u'^2
+\frac{1}{12}u''\big)e^u
\nonumber \\
&&+\lambda^4\big(\frac{1}{360}u''''+\frac{1}{180}u'u'''
+\frac{7}{1440}u'^2u''+\frac{1}{1920}u'^4+\frac{1}{240}u''^2\big)e^u\Big)'~,
\nonumber \\
&&~~~~~~~~
\label{eq:14}
\\
\frac{\partial u}{\partial t_{1,P}} &=&
\Big(uv+\lambda^2\frac{1}{6}v''+\lambda^4
\frac{-1}{360}v''''\Big)'~,
\\
\frac{\partial v}{\partial t_{1,P}} &=& \Big(\frac{1}{2}v^2+(u-1)e^u+\lambda^2
\big(\frac{1}{24}(u+3)u'^2+\frac{1}{12}(u+2)u''\big)e^u
\nonumber \\
&&+\lambda^4\big(\frac{1}{360}(u+4)u''''
+\frac{1}{180}(u+5)u'u'''+\frac{7}{1440}(u+6)u''u'^2
\nonumber \\
&&~~~~~~+\frac{1}{1920}(u+7)u'^4+\frac{1}{240}(u+5)u''^2\big)e^u\Big)'~,
\\
\frac{\partial u}{\partial t_{1,Q}} &=& \Big(\frac{1}{2} v^2+e^u
+\frac{\lambda^2}{24}(3u'^2+4u'')e^u
\nonumber \\
&&+\lambda^4 \big( \frac{1}{120}u''''+\frac{1}{48}u'u'''
+\frac{1}{48}u'^2u''+\frac{1}{384}u'^4+\frac{1}{72}u''^2 \big) e^u\Big)' ~,
\\
\frac{\partial v}{\partial t_{1,Q}} &=& \Big(ve^u
+\frac{\lambda^2}{24}(u'^2v+2u'v'+2u''v+4v'')e^u
\nonumber \\
&&+\lambda^4 \big( \frac{1}{360}vu''''+\frac{1}{180}vu'u'''
+\frac{7}{1440}vu'^2u''+\frac{1}{1920}vu'^4
\nonumber \\
&&~~~~~+\frac{1}{240}vu''^2
+\frac{1}{180}v'u'''+\frac{7}{720}v'u'u''+\frac{1}{480}v'u'^3
\nonumber \\
&&~~~~~+\frac{11}{720}v''u''+\frac{1}{120}v''u'^2+\frac{1}{80}v'''u'
+\frac{1}{120}v'''' \big) e^u\Big)'~.
\label{eq:15}
\end{eqnarray}
Here $\lambda^2$ is the genus expansion parameter. In deriving the above
equations we first write down candidate correction terms in the
right-hand-side of (\ref{eq:14})-(\ref{eq:15}) with unknown coefficients
and then fix them by
demanding the commutativity of the flows, e.g. the vanishing of the
cross-derivatives such as
$\partial /\partial t_{1,P}(\partial v/\partial t_{0,Q})-
\partial/\partial t_{0,Q}(\partial v/\partial t_{1,P})=0$. The flow
commutativity
leads to an over-determined system for the unknown coefficients and gives
rise to their unique solution.

We now turn to the construction of
a matrix model which reproduces
the $CP^1$ model.
Let us first recall some basic formulas for the analysis of matrix
models using orthogonal polynomials \cite{Da}.
We consider a one-matrix model with an action ${\rm Tr}\, V(M)$ which
depends only on the eigenvalues of the $N\times N$ hermitian matrix $M$.
The orthogonal polynomials are defined by
\begin{eqnarray}
&&\int d\lambda e^{NV(\lambda)}\psi_n(\lambda)\psi_m(\lambda)=\delta_{n,m}h_m,
\\
&&~~~~~~\psi_n(\lambda)=\lambda^n+\mbox{lower terms}~.
\end{eqnarray}
The multiplication by $\lambda$ of the orthogonal polynomials leads to a
three-term recursion relation
\begin{eqnarray}
&& \lambda\psi_n(\lambda) = \sum_m Q_{nm}\psi_m(\lambda)=
\psi_{n+1}(\lambda)+v_n\psi_n(\lambda)
+R_n\psi_{n-1}(\lambda)~,
\\
&& \hskip10mm R_n = h_n/h_{n-1}\equiv e^{N(\phi_n-\phi_{n-1})}.
\end{eqnarray}
The free-energy of the system is given by $F=\log \Pi_0^{N-1} h_n
=N\sum_0^{N-1}\phi_n$.
The matrix $Q$ has non-vanishing elements only on the diagonal and
1st off-diagonal lines and thus is a Jacobi matrix.
In the continuum limit $N(\phi_n-\phi_{n-1})\approx F''/N^2=u$
and the matrix $Q$ turns into the Lax operator
\begin{equation}
Q\approx L=p+v+e^u p^{-1}~.
\end{equation}
$Q$ replaces $L$ in the discussion of the $CP^1$ model at higher genera.
Time evolution equations are now given by the commutation relations
\begin{eqnarray}
&&\frac{n}{N}\frac{\partial Q}{\partial t_{n-1,Q}}=[(Q^n)_+,Q]~,
{}~~~~~~~~~~~~~~~n=1,2,\cdots
\label{eq:16}
\\
&&\frac{1}{2N}\frac{\partial Q}{\partial t_{n,P}}=[(Q^n(\log Q-c_n))_+,Q]~,
{}~~n=0,1,2,\cdots,
\label{eq:17}
\end{eqnarray}
where $+$ means to take the upper-triangular part of a matrix including the
diagonal line.
It is easy to check that the matrix commutators reduce to the Poisson bracket
(17) in the continuum limit and thus (\ref{eq:16}) (\ref{eq:17})
reduce to (\ref{eq:12}) (\ref{eq:13}) at
genus $g=0$.

The derivative operator $d/d\lambda$, on the other hand,
becomes a lower- \\
triangular matrix when acting on the orthogonal polynomials
\begin{equation}
\frac{d}{d\lambda}\psi_n(\lambda)=n\psi_{n-1}(\lambda)+\cdots
=\sum_m P_{nm}\psi_m(\lambda)~.
\end{equation}
By partial integration we have
\begin{eqnarray}
\int d\lambda e^{NV(\lambda)}\frac{d}{d\lambda}\psi_n(\lambda)\psi_m(\lambda)
&=& -\int d\lambda e^{NV(\lambda)}NV'(\lambda)\psi_n(\lambda)\psi_m(\lambda)
\nonumber \\
&=& -NV'(Q)_{nm}h_m~,~~~~~~~~n>m.
\end{eqnarray}
Hence
\begin{equation}
P=-NV'(Q)_-~,
\end{equation}
where $-$ means to take the lower-triangular part of the matrix.
The string equation then reads
\begin{equation}
[Q,P]=-N[Q,V'(Q)_-]=1~.
\end{equation}

Let us now introduce our matrix model defined by an action
\begin{eqnarray}
{\rm Tr}\, V(M) &=& -2{\rm Tr}\, M(\log M-1)
+\sum_{n=1}2t_{n,P}{\rm Tr}\, M^n(\log M-c_n)
\nonumber \\
&&+\sum_{n=1} \frac{1}{n}t_{n-1,Q}{\rm Tr}\, M^n~.
\end{eqnarray}
We then obtain
\begin{eqnarray}
&&N[V'(Q)_-,Q] \nonumber \\
&=&-2N[(\log Q)_-,Q]
+N\sum 2nt_{n,P}[(Q^{n-1}(\log Q-c_{n-1}))_-,Q]
\nonumber \\
&&+N\sum t_{n-1,Q}[(Q^{n-1})_-,Q]
\nonumber \\
&=& \frac{\partial Q}{\partial t_{0,P}}
-\sum_{n=1}nt_{n,P}\frac{\partial Q}{\partial t_{n-1,P}}
-\sum_{n=2}(n-1)t_{n-1,Q}\frac{\partial Q}{\partial t_{n-2,Q}}~,
\label{eq:19}
\end{eqnarray}
where we used (\ref{eq:16}),(\ref{eq:17}) and the ``scaling'' property of
our logarithmic potentials (\ref{eq:18}).
(46) is now compactly expressed as
\begin{equation}
1+\sum_{n=1}\sum_{\alpha} nt_{n,\alpha}
\frac{\partial Q}{\partial t_{n-1,\alpha}}
=\frac{\partial Q}{\partial t_{0,P}}~,~~~~~~~~(\alpha=P,Q)~.
\label{eq:20}
\end{equation}
The diagonal element of (\ref{eq:20})
reads
\begin{equation}
\frac{\partial \langle PQ \rangle}{\partial t_{0,P}}
=1+\sum_n \sum_{\alpha}nt_{n,\alpha}
\langle\sigma_{n-1}(\Phi_{\alpha})PQ\rangle
\label{eq:21}
\end{equation}
in terms of the correlation functions. By integrating (\ref{eq:21}) we obtain
\begin{eqnarray}
&&u=t_{0,Q}+\sum_n \sum_{\alpha}nt_{n,\alpha}
\langle\sigma_{n-1}(\Phi_{\alpha})P\rangle~,
\\
&&v=t_{0,P}+\sum_n \sum_{\alpha}nt_{n,\alpha}
\langle\sigma_{n-1}(\Phi_{\alpha})Q\rangle~.
\end{eqnarray}
These are the string equations of the $CP^1$ model \cite{DW}.

Therefore, in order to demonstrate the realization of the $CP^1$
model by our matrix integral we finally have to examine
our flow equations (\ref{eq:16}),(\ref{eq:17}) and check if
they agree with the flow equations of the $CP^1$ model at higher genera.
Checking (\ref{eq:16}) is straightforward.
In the cases $n=1,2$, for instance, (\ref{eq:16}) gives
\begin{eqnarray}
&&\frac{\partial \phi_n}{\partial t_{0,Q}}=v_n~,
\\
&&\frac{\partial v_n}{\partial t_{0,Q}}=N(R_{n+1}-R_n)~,
\label{eq:22}
\\
&&\frac{\partial \phi_n}{\partial t_{1,Q}}=\frac{1}{2}(R_{n+1}+R_n+v_n^2)~,
\\
&&\frac{\partial v_n}{\partial t_{1,Q}}=\frac{N}{2}((v_{n+1}+v_n)R_{n+1}-
(v_n+v_{n-1})R_n)~.
\end{eqnarray}
If we use the definition $t_{0,P}=n/N,u_n=\phi_n',
R_n=\exp\big(N(\phi_n-\phi_{n-1})\big)$
and
Taylor-expand $R_{n\pm 1}
(v_{n\pm 1})$ around $R_n (v_n)$,
we easily recover the $CP^1$ equations.
For instance, by expanding the right-hand-side of (\ref{eq:22}) as
\begin{eqnarray}
N(R_{n+1}-R_n) &=& e^{u_n} \Big( u_n'+\frac{u_n'''}{12N^2}+\frac{u_n'^3}{24N^2}
+\frac{u_n'u_n''}{6N^2}
\nonumber \\
&&+\frac{u_n'^5}{1920N^4}+\frac{u_n'^3u_n''}{144N^4}+\frac{u_n'u_n''^2}{72N^4}
+\frac{u_n'^2u_n'''}{96N^4}
\nonumber \\
&&+\frac{u_n''u_n'''}{72N^4}+\frac{u_n'u_n''''}{120N^4}
+\frac{u_n'''''}{360N^4}+\cdots \Big)
\end{eqnarray}
we reproduce (\ref{eq:14}). Note that the genus expansion parameter
$\lambda^2$ becomes $1/N^2$.

Checking (\ref{eq:17}), on the other hand, is rather non-trivial. We first
have to
compute the matrix elements of the logarithm of a Jacobi matrix $Q$.
We use a formula,
\begin{equation}
(\log Q)_{nm}=\frac{1}{2\pi i}\oint \frac{dz}{z}z^{n-m}
\log\Big(z+z^{-1}R(n+z\frac{d}{dz})+v(n+z\frac{d}{dz})\Big)\cdot \bf{1}
\end{equation}
and expand the arguments of $R$ and $v$. We can then compute $\log Q$
perturbatively
for each fixed order of differentiations of $R,v$. Up to the 2nd order
in the derivatives, for instance, we obtain
\begin{eqnarray}
(\log Q)_{n,n} &=& \frac{1}{2}\log R_n+\frac{1}{4N}\frac{R'_n}{R_n}
+\frac{1}{24N^2}\frac{R_n''}{R_n}-\frac{1}{24N^2}\frac{R_n'^2}{R_n^2}~,
\\
(\log Q)_{n,n-1} &=& \frac{1}{2}v_n-\frac{1}{4N}v_n'+\frac{1}{24N^2}v_n''~,
\\
(\log Q)_{n,n+1} &=& \frac{1}{2}\frac{v_n}{R_n}
+\frac{1}{4N} \Big( \frac{v_n'}{R_n}-
\frac{2R_n'v_n}{R_n^2} \Big)+\frac{1}{24N^2}\frac{v_n''}{R_n}
\nonumber \\
&&-\frac{1}{4N^2}\frac{R_n''v_n}{R_n^2}
+\frac{1}{2N^2}\frac{R_n'^2v_n}{R_n^3}-\frac{1}{4N^2}\frac{R_n'v_n'}{R_n^2}~.
\end{eqnarray}
Using the explicit expressions for $\log Q$ as above one can evaluate the
right-hand-side of (\ref{eq:17}) and check its agreement with the $CP^1$
equations. We have calculated $\log Q$ up to the 3rd order in the derivatives
and have confirmed the agreement of (\ref{eq:17}) with the $t_{n,P}$ flow
equations up to $n=2$ and $g=1$.

We should note that in our construction we take the ordinary large-$N$ limit
of the matrix model rather than the double scaling limit: we do not adjust the
coupling constants to any specific set of values. This is in accord with the
idea that the $CP^1$ model, due to its asymptotic freedom, has no phase
transition points. In this sense our model is in a spirit closer to the
Kontsevich \cite{K} or Penner \cite{P} model rather than the double-scaled
matrix
models describing the $c<1$ minimal matter coupled to the two-dimensional
gravity \cite{BK}. We think it is fortunate that the $CP^1$ model has
a relatively simple representation in terms of a single matrix integral.
Our model may prove useful in the study of the moduli space of holomorphic
maps from  Riemann surface to $CP^1$. It will be very interesting
to see if our
construction may be generalized to $CP^n$ or Grassmann manifolds.

\bigskip

T.E. would like to thank E. Witten for a conversation on the subject of this
paper.
The research of T.E. and S.K.Y. is supported in part by Grant-in-Aid
for Scientific Research on Priority Area 231 ``Infinite Analysis'',
Japan Ministry of Education.

\end{document}